\documentclass[epj]{svjour}

\usepackage{graphicx}
\usepackage{fancyhdr}
\def\makeheadbox{{
 }}
\def\mytitle{Indirect dark matter searches with H.E.S.S.}
\def\myauthors{Moulin}
\def\mytype{Contributed Talk}
\def\mysession{Cosmology and Astrophysics}

\begin{document}
\title{Indirect dark matter searches with H.E.S.S.}
\subtitle{}
\author{E. Moulin\inst{1}
\thanks{\emph{Email:} emmanuel.moulin@cea.fr}%
\,\,on behalf of the H.E.S.S. collaboration\inst{2}
}                     
%
%
\institute{DAPNIA/DSM/CEA, CE Saclay F-91191 Gif-sur-Yvette,
Cedex, France \and http://www.mpi-hd.mpg.de/hfm/HESS/}
%
\date{}
\abstract{The annihilations of WIMPs produce high energy
gamma-rays in the final state. These high energy gamma-rays may be
detected by IACTs such as the H.E.S.S. array of Imaging
Atmospheric Cherenkov telescopes. Besides the popular targets such
as the Galactic Center or galaxy clusters such as VIRGO, dwarf
spheroidal galaxies are privileged targets for Dark Matter
annihilation signal searches. H.E.S.S. observations on the
Sagittarius dwarf galaxy are presented. The modelling of the Dark
Matter halo profile of Sagittarius dwarf is discussed. Constraints
on the velocity-weighted cross section of Dark Matter particles
are derived in the framework of Supersymmetric and Kaluza-Klein
models. The future of H.E.S.S. will be briefly discussed.
} 
\maketitle
%

\section{Introduction}
\label{intro} H.E.S.S. (High Energy Stereoscopic System) is an
array of four Imaging Atmospheric Cherenkov Telescopes (IACTs)
located in the Khomas Highlands of Namibia at an altitude of 1800
m above sea level. Each telescope has an optical reflector of 107
m$^2$ composed of 382 round mirrors \cite{bernlohr}. The Cherenkov
light is emitted by charged particles from electromagnetic showers
initiated by the interaction of the primary gamma-rays in the
Earth's upper atmosphere. The light is collected by the reflector
which focuses it on a camera made of 960 fast photomultiplier
tubes (PMTs) with individual field of view of 0.16$^{\circ}$ in
diameter \cite{vincent}. Each PMT is equipped with a Winston cone
to maximize the light collection and limit the background light.
The total field of view of the H.E.S.S. instrument is 5$^{\circ}$
in diameter. The stereoscopic technique allows for an accurate
reconstruction of the direction and energy of the primary
gamma-rays as well as for an efficient rejection of the background
induced by cosmic ray interactions. The energy threshold is about
100 GeV at zenith and the angular resolution is better than
0.1$^{\circ}$ per gamma-ray. The point source sensitivity is $\rm
2 \times 10^{-13} cm^{-2}s^{-1}$
above 1 TeV for a 5$\sigma$ detection in 25 hours \cite{crabe}.\\
The H.E.S.S. instrument is designed to detect very high energy
(VHE) gamma-rays in the 100 GeV - 100 TeV energy regime and
investigate their possible origin. Scenarios beyond the Standard
Model of particle physics predict plausible WIMP (weakly
interacting massive particle) candidates to account for the Cold
Dark Matter. Among these are the minimal supersymmetric extension
of the Standard Model (MSSM) or universal extra dimension (UED)
theories. With R-parity conservation, SUSY models predict the
lightest SUSY particle (LSP) to be stable. In various SUSY
breaking scenarios, the LSP is the lightest neutralino $\rm
\tilde{\chi}$. In Kaluza-Klein (KK) models with KK-parity
conservation, the lightest KK particle (LKP) is stable \cite{KK},
the most promising being the first KK mode of the hypercharge
gauge boson, $\rm \tilde{B}^{(1)}$. The annihilation of WIMPs in
galactic halos may produce gamma-ray signals detectable with
Cherenkov telescopes (for a review, see \cite{bertone}).
Generally, their annihilations will lead to a continuum of
gamma-rays with energy up to the WIMP mass resulting from the
hadronization and decay of the cascading products, mainly from
$\pi^0$'s generated in the quark jets. The gamma-ray flux from DM
particle annihilations of mass m$_{DM}$ in a spherical halo can be
expressed as:
\begin{equation}
\frac{d\Phi(\Delta\Omega,E_{\gamma})}{dE_{\gamma}} =
\frac{1}{4\pi}\,\frac{\langle \sigma v
\rangle}{m^2_{DM}}\frac{dN_{\gamma}}{dE_{\gamma}}\,\times\,\bar{J}\Delta\Omega
\end{equation}
which is a product of a particle physics term containing the
velocity-weighted cross section $\langle \sigma v \rangle$ and the
differential gamma-ray spectrum $dN_{\gamma}/dE_{\gamma}$, and an
astrophysics term $\bar{J}$ given by:
\begin{equation}
\bar{J} =
\frac{1}{\Delta\Omega}\int_{\Delta\Omega}\int_{l.o.s}\rho^2 ds
\end{equation}
This term corresponds to the line-of-sight-integrated squared
density of the DM distribution which is averaged over the solid
angle integration region spanned by the H.E.S.S. point spread
function (PSF).\\
Plausible astrophysical targets have been proposed to search for
DM, from local galactic objects to extragalactic objects such as
galaxy clusters. This paper reports on three recent results on
targets relevant to indirect dark matter search: the Galactic
Center, the center of the VIRGO cluster (M87) and the Sagittarius
dwarf spheroidal galaxy.

\section{The Galactic Center}
\label{sec:1} H.E.S.S. observations towards the Galactic Center
have revealed in 2004 a bright point-like gamma-ray source, HESS
J1745-290 \cite{GC}, coincident in position with the supermassive
black hole Sgr A$^*$, with a size lower than 15 pc. Diffuse
emission along the Galactic plane has also been detected
\cite{diffuse} and correlates well with the mass density of
molecular clouds from the Central Molecular Zone, as traced by CS
emission \cite{CS}. According to recent detailed studies
\cite{GC_position}, the source position is located at an angular
distance of $\rm 7.3'' \pm 8.7'' (stat.) \pm 8.5'' (syst.)$ from
Sgr A$^*$. The pointing accuracy allows to discard the association
of the VHE emission with the center of the radio emission of the
supernov\ae \ remnant Sgr A East but the association with the
pulsar wind nebula
G359.95-0.04 can not be ruled out.\\
From 2004 data set, the energy spectrum of the source is well
fitted in the energy range 160 GeV - 30 TeV to a power-law
spectrum $\rm dN/dE \propto E^{-\Gamma}$ with a spectral index
$\rm \Gamma = 2.25 \pm 0.04\,(stat.) \pm 0.1\,(syst.)$. No
deviation from a power-law is observed leading to an upper limit
on the energy cut-off of 9 TeV (95\% C.L.). The VHE emission from
HESS J1745-290 does not show any significant periodicity or
variability from 10 minutes
to 1 year \cite{GC_periodicity}.\\
Besides plausible astrophysical origins (see e.g. \cite{GC_astro}
and references therein), an alternative explanation is the
annihilation of DM in the central cusp of our Galaxy \cite{GC_DM}.
The spectrum of HESS J1745-290 shows no indication for gamma-ray
lines.
\begin{figure}[!b]
\includegraphics[width=0.45\textwidth,height=0.3\textwidth,angle=0]{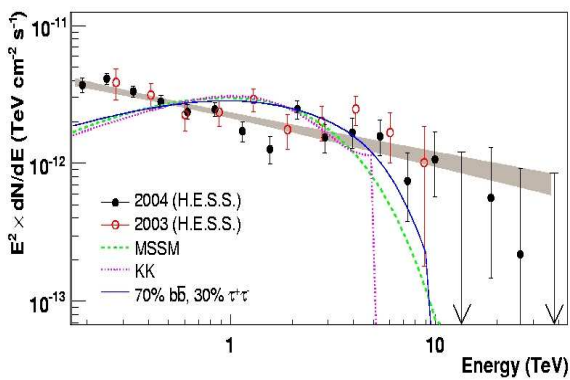}
\vspace{0.2cm} \caption{Spectral energy density $\rm E^2 \times
dN/dE$ of $\gamma$-rays from  HESS J1745-290 for 2003 (red empty
circles) and 2004 (black filled circles) datasets of the H.E.S.S.
observation of the Galactic Center. The shaded area shows the best
power-law fit to the 2004 data points. The spectra expected from
the annihilation of a MSSM-like 14 TeV neutralino (dashed green
line), a 5 TeV KK DM particle (dotted pink line) and a 10 TeV DM
particle annihilating into 70\% b $\rm\bar{b}$ and 30\%
$\rm\tau^+\tau^-$ in final state (solid blue line) are presented.}
\label{fig:1}       
\end{figure}
\begin{figure}
\includegraphics[width=0.5\textwidth,height=0.4\textwidth,angle=0]{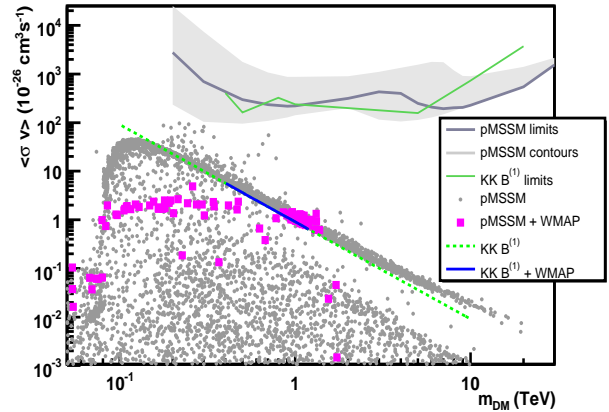}
\caption{Exclusion limit for the Galactic Center on $\rm \langle
\sigma v\rangle$ versus the WIMP mass for a NFW profile. The
limits are derived from H.E.S.S. data in the case of pMSSM and KK
dark matter models. The predictions for pMSSM (grey points) and KK
(green dashed line) models are plotted. Those satisfying in
addition the WMAP constraints on the relic cold dark matter
density are superimposed for pMSSM (pink boxes) and KK (blue solid
segment) models.}
\label{fig:2}       
\end{figure}
The observed $\gamma$-ray flux may result from secondaries of DM
annihilation such as $\rm b\bar{b}$, etc. When $\rm E^2d\Phi/dE$
is plotted as a function of the $\gamma$-ray energy E$_{\gamma}$,
the WIMP annihilation is characterized by a plateau for
E$_{\gamma}$ much lower than the WIMP mass m$_{DM}$. A rapid
decrease is observed when $\rm E_{\gamma}\sim m_{DM}$. The
spectrum extends up to masses of about 10 TeV which requires large
neutralino masses ($>$10 TeV). They are unnatural in
phenomenological MSSM scenarios. The Kaluza-Klein models provide
harder spectra which still significantly deviate from the measured
one. Non minimal version of the MSSM may yield flatter spectrum
with mixed 70\% b $\rm\bar{b}$ and 30\% $\rm\tau^+\tau^-$ final
states \cite{profumo}. Even this scenario does not fit to the
measured spectrum. The hypothesis that the spectrum measured by
H.E.S.S. originates only from DM particle annihilations is highly
disfavored. The observed signal may be produced by the
superposition of a DM annihilation signal and a power-law
astrophysical background spectrum. In this case, using a NFW
profile, constraints on the velocity-weighted annihilation
cross-section $\rm \langle \sigma v \rangle$ can be derived.
Fig.~\ref{fig:2} shows the 99\% confidence level limits on $\rm
\langle \sigma v \rangle$ from H.E.S.S. data in the case of pMSSM
and KK DM models. Predictions from pMSSM models calculated with
DarkSUSY4.1 \cite{darksusy} and KK models from \cite{KK} are shown
as well as those satisfying cosmological constraints. The limits
on $\rm \langle \sigma v \rangle$ derived from H.E.S.S. data for
both DM models are $\mathcal{O}\rm(10^{-24}) cm^3s^{-1}$. With a
NFW profile, no $\rm \langle \sigma v \rangle$ from SUSY or KK DM
models can be ruled out.

\section{The VIRGO galaxy cluster}
\label{sec:2}
 Astrophysical systems such as the VIRGO galaxy cluster
\cite{virgo} have also been considered as targets for DM
annihilations. The elliptical galaxy M87 at the center of the
VIRGO cluster has been observed by H.E.S.S. It is an active galaxy
located at 16.3 Mpc which hosts a black hole of $\rm 3.2 \times
10^9 M_{\odot}$. From 2003 to 2006, H.E.S.S. detected a 13$\sigma$
signal in 89 hour observation time at a location compatible with
the nominal position of the nucleus of M87 \cite{M87}. With the
angular resolution of H.E.S.S., the VHE source is point-like, with
an upper limit on the size of 3' at 99\% confidence level.
\begin{figure}[!t]
\includegraphics[width=0.45\textwidth,height=0.4\textwidth,angle=0]{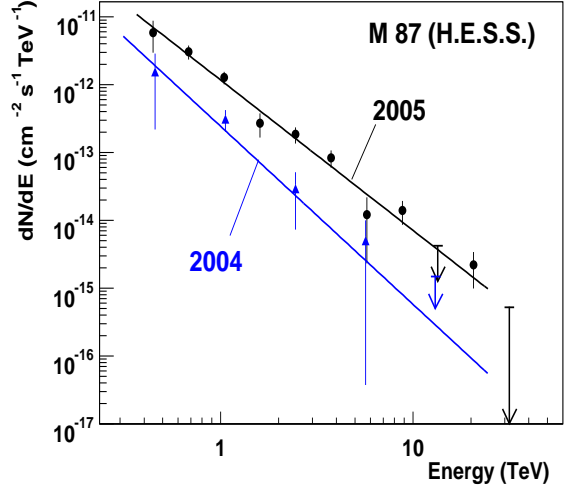}
\vspace{0.2cm} \caption{Differential energy spectrum of M87 from
$\sim$~400 GeV up to 10 TeV from 2004 and 2005 data. Corresponding
fits to a power-law function  $\rm dN/dE \propto E^{-\Gamma}$ are
plotted as solid lines. The data exhibit indices of $\rm \Gamma =
2.62 \pm 0.35$ for 2004 and $\rm \Gamma = 2.22 \pm 0.15$ for
2005.}
\label{fig:3}       
\end{figure}Given
the distance of M87, this corresponds to a size of 13.7 kpc.\\
Fig.~\ref{fig:3} shows the energy spectrum for 2004 and 2005 data.
Both data sets are well fitted to a power-law spectrum with
spectral indices $\Gamma = 2.62 \pm 0.35$ (2004) and $\Gamma =
2.22 \pm 0.25$ (2005). The 2005 average flux is found to be higher
by a factor $\sim$5 the the average flux of 2004. The integrated
flux above 730 GeV shows a yearly variability at the level of
3.2$\sigma$ \cite{M87}. Variability at the day time scale has been
observed in the 2005 data at the level of 4$\sigma$. This fast
variability puts stringent constraints on the size of the VHE
emission region. The position centered on M87 nucleus excludes the
center of the VIRGO cluster and outer radio regions of M87 as the
gamma-ray emission region. The observed variability at the scale
of $\sim$ 2 days requires a compact emission region below 50 R$_s$
where $\rm R_s \sim 10^{-15}$ cm is the Schwarzschild
radius of the M87 supermassive black hole \cite{M87}.\\
The short and long term temporal variability observed with
H.E.S.S. excludes the bulk of the TeV gamma-ray signal to be of DM
origin.

\section{The Sagittarius dwarf spheroidal galaxy}
\label{sec:3} 
\begin{figure}[!t]
\includegraphics[width=0.5\textwidth,height=0.45\textwidth,angle=0]{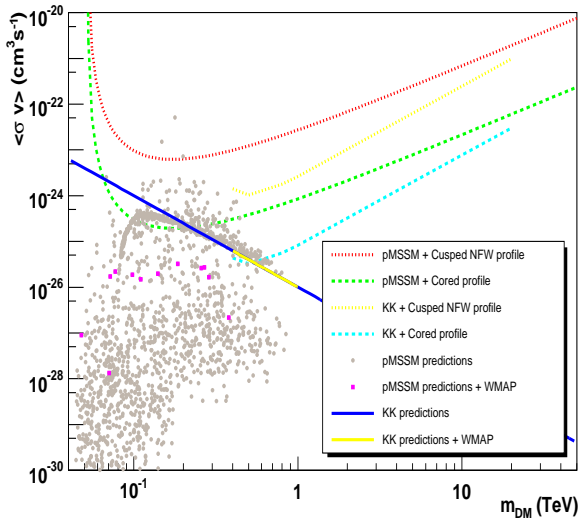}
\caption{Upper limits at 95\% C.L. on $\rm \langle \sigma
v\rangle$ versus the dark matter particle mass in the case of a
NFW and cored DM halo profile for the Sagittarius dwarf galaxy.
The limits are derived from H.E.S.S. data in the case of pMSSM and
KK dark matter models. The predictions in pMSSM and KK models are
also plotted with in addition those satisfying in addition the
WMAP constraints on the relic cold dark matter density.}
\label{fig:4}       
\end{figure}

Nearby dwarf spheroidal galaxies are known to be
among the best targets to search for DM signals, considered as the
most extreme DM-dominated environments. Unlike the Galactic
Center, those galaxies are expected to have reduced astrophysical
backgrounds. The Sagittarius (Sgr) dwarf galaxy is a Galaxy
satellite of the Local Group. It was recently discovered
\cite{ibata} in the direction of the Galactic Center. The center
of Sgr is located at 24~kpc from the Sun. The visible mass profile
is difficult to measure because of the contamination of foreground
galactic stars. The central velocity dispersion has been measured
by various groups \cite{zaggia}.\\
Sagittarius dwarf has been observed in 2006 by H.E.S.S. with an
averaged zenith angle of 19$^{\circ}$. A total of 11 hours of high
quality data is available after standard selection cuts. The
target position is chosen at the location of the Sgr luminous
cusp, coincident with the center of the globular cluster M54
\cite{M54}. The pointing position is thus $\rm (RA =
18^h55^m59.9s, Dec = -30d28'59.9'')$ in equatorial coordinates
(J2000.0). The annihilation signal from Sgr is expected to come
from a region of 1.5 pc, thus much smaller than the H.E.S.S. PSF
so that a point-like signal has been searched for. No significant
gamma-ray excess is detected at the nominal target position. We
thus derived an upper limit at 95\% C.L. on the observed number of
gamma-rays using the Feldman \& Cousins method \cite{upperlimit}.
A 95\% confidence level upper limit on the gamma-ray flux is also
derived :
\begin{equation}
\rm \Phi_{\gamma}(E_{\gamma} > 250 GeV) < 3.6 \times 10^{-12}
cm^{-2}s^{-1}
\end{equation}
assuming a power-law spectrum of spectral index $\Gamma = 2.2$. In
order to constraint on the velocity-weighted annihilation cross
section, a modelling of the Sgr DM halo has been carried out.
Dynamic models have been studied in \cite{evans}. Two plausible
models of the mass distribution of the DM halo have been studied:
a NFW profile used as a reference model and a cored isothermal
profile. For the NFW profile, the structure parameters are
extracted from \cite{evans}. In case of the cored model, the
velocity dispersion is assumed to be independent of the position
and equal to the central velocity dispersion $\sigma = 8.2\,\pm\,
0.3\,\rm km\,s^{-1}$ as reported in \cite{zaggia}. The velocity
dispersion tensor is taken isotropic as in \cite{evans}. The mass
density profile of stars is extracted from \cite{M54}. The cored
profile has a small core radius due to a cusp in the luminous
profile. The value of $\rm \bar{J}$ is found to be larger for the
cored profile than for the NFW
profile \cite{J_sgr}.\\
Fig.~\ref{fig:4} presents the constraints on the velocity-weighted
annihilation cross section in the case of a cored (green dashed
line) and cusped NFW (red dotted line) profiles in the solid angle
integration region $\rm \Delta\Omega = 2\times 10^{-5}$sr.
Predictions for phenomenological MSSM (pMSSM) \\models are
displayed (grey points) as well as those fulfilling in addition
the WMAP constraints on the CDM relic density (blue points). The
pMSSM
models are calculated with DarkSUSY4.1~\cite{darksusy}. In the
case of a cusped NFW profile, the H.E.S.S. observations do not set
severe constraints on the velocity-weighted cross section. For a
cored profile, due to a higher central density, stronger
constraints are derived and some pMSSM models can be excluded in
the upper part of the pMSSM scanned region. In the case of KK dark
matter, the differential gamma-ray spectrum is parametrized using
Pythia \cite{pythia} simulations and branching ratios from
\cite{KK}. Predictions for the velocity-weighted annihilation
cross section of B$^{(1)}$ dark matter particle are performed
using the formula given in~\cite{sigmaKK}. In this case, the
expression for $\langle \sigma v \rangle$ depends analytically on
the $\rm \tilde{B^{(1)}}$ mass square. The KK models are also
shown (solid blue line) as well as those satisfying the WMAP
constraint on the cold dark matter density (solid yellow line).
The sensitivity of H.E.S.S. in the case of Kaluza-Klein models
where the hypercharge boson B$^{(1)}$ is the LKP, is also
displayed on Fig.~\ref{fig:4} for a cored (light blue solid line)
and a cusped NFW (yellow solid line) profile respectively using
the value of $\rm \bar{J}$ computed in section 3.2. With a NFW
profile, no Kaluza-Klein models can be tested. In the case of a
cored profile, some models providing a LKP relic density
compatible with WMAP contraints can be excluded. From the
sensitivity of H.E.S.S., we exclude B$^{(1)}$ masses lying in the
range 300 - 500 GeV.

\section{Conclusion}
\label{conclusion} The H.E.S.S. collaboration has studied
potential targets for DM annihilations. The TeV $\gamma$-ray
energy spectrum measured by H.E.S.S. in the Galactic Center region
is unlikely to be interpreted in terms of WIMP annihilations.
Constraints on the velocity-weighted annihilation cross-section
have been derived in the case of a NFW DM halo profile. Neither
pMSSM nor KK models can be ruled out. The detection of a temporal
variability of the VHE signal from the M87 nucleus in the VIRGO
galaxy cluster excludes the bulk of the gamma-ray signal to come
from DM annihilations. H.E.S.S. observed the Sagittarius dwarf
spheroidal galaxy and no significant gamma-ray excess has been
detected. Two DM halo modellings of Sgr have been investigated to
derive contraints on the velocity-weighted annihilation
cross-section.
Some models can be ruled out in the case of a cored profile.\\
Dark matter searches will continue and searches with H.E.S.S. 2
will start in 2009.  The phase 2 will consist of a new large 28 m
diameter telescope located at the center of the existing array.
With the availability of the large central telescope H.E.S.S. 2,
the analysis energy threshold will be lowered down to less than 80
GeV and will allow to explore more supersymmetric models.


\end{document}